\documentclass[article]{aastex63}
\pdfoutput=1

\usepackage{amssymb,amsmath,graphicx,multirow}
 
\shorttitle{RSGs in NGC 6822}
\shortauthors{Dimitrova et al.}

\begin{document}
\title{Locating Red Supergiants in the Galaxy NGC 6822}

\author[0000-0002-4319-1615]{Tzvetelina A. Dimitrova}
\affiliation{Earth and Space Exploration, Arizona State University, P.O. Box 871404, Tempe, AZ 85287-1404, USA}
\affiliation{Department of Astronomy, University of Washington, 3910 15th Ave NE, Seattle, WA 98105, USA}

\author[0000-0002-5787-138X]{Kathryn F. Neugent}
\affiliation{Department of Astronomy, University of Washington, 3910 15th Ave NE, Seattle, WA 98105, USA}
\affiliation{Dunlap Institute for Astronomy \& Astrophysics, University of Toronto, 50 St. George Street, Toronto, ON M5S 3H4, Canada}
\affiliation{Lowell Observatory, 1400 W Mars Hill Road, Flagstaff, AZ 86001, USA}

\author[0000-0001-6563-7828]{Philip Massey}
\affiliation{Lowell Observatory, 1400 W Mars Hill Road, Flagstaff, AZ 86001, USA}
\affiliation{Department of Astronomy and Planetary Sciences, Northern Arizona University, PO Box 6010, Flagstaff, AZ 86011-6010, USA}

\author[0000-0003-2184-1581]{Emily M. Levesque}
\affiliation{Department of Astronomy, University of Washington, 3910 15th Ave NE, Seattle, WA 98105, USA}

\begin{abstract}
Using archival near-IR photometry, we identify 51 of the K-band brightest red supergiants (RSGs) in NGC 6822 and compare their physical properties with stellar evolutionary model predictions. We first use Gaia parallax and proper motion values to filter out foreground Galactic red dwarfs before constructing a $J-K$ vs. $K$ color-magnitude diagram to eliminate lower-mass asymptotic giant branch star contaminants in NGC 6822. We then cross-match our results to previously spectroscopically confirmed RSGs and other NGC 6822 content studies and discuss our overall completeness, concluding that radial velocity alone is an insufficient method of determining membership in NGC 6822. After transforming the $J$ and $K$ magnitudes to effective temperatures and luminosities, we compare these physical properties with predictions from both the Geneva single-star and Binary Population and Spectral Synthesis (BPASS) single and binary star evolution tracks. We find that our derived temperatures and luminosities match the evolutionary model predictions well, however the BPASS model that includes the effects of binary evolution provides the best overall fit. This revealed the presence of a group of cool RSGs in NGC 6822, suggesting a history of binary interaction. We hope this work will lead to further comparative RSG studies in other Local Group galaxies, opportunities for direct spectroscopic follow-up, and a better understanding of evolutionary model predictions.
\end{abstract}

\section{Introduction} \label{sec:intro}
Red supergiants (RSGs) are the coolest of the evolved massive stars, with K and M spectral types and temperatures below $4500$K. They start their lives on the main sequence, spending $10^6-10^7$ years as hydrogen fusing OB-type stars with initial masses of 8-30$M_\odot$. Once the core hydrogen supply of these progenitor stars is depleted, they dramatically expand in size over a short period of time as they leave the main sequence and move across the HR-diagram, evolving into a phase of helium core fusion. After briefly passing through a yellow supergiant (YSG) phase, the expansion increases their radius up to a thousand times that of the Sun, and they are now cool RSGs with effective temperatures of $\sim$3500-4500 K. RSGs continue on to fuse heavier elements in their cores before reaching the end of their life cycles with a core collapse death that leaves behind either neutron stars or black holes (see \citealt{EmilyRSGBook} and references therein for an overview of RSG evolution).

A number of recent studies have been made of the RSG content of nearby galaxies, thanks both to Gaia and improved near-IR (NIR) surveys. In the last few years, both \citet{Massey2021} and \citet{Ren2021} used this method to identify RSGs in the high metallicity spiral galaxies of M31 and M33. The Large and Small Magellanic Clouds (LMC, SMC) provide low-metallicity complements in the Local Group and their RSG contents were recently determined by \citet{YangLMC} and \citet{LMCBinFrac} for the LMC and \citet{YangSMC} and \citet{WRRSG} for the SMC. Additionally, \citet{Dimitrova2020} located 138 RSGs in the starburst galaxy IC 10 following a similar process. Here we extend these RSG content studies to the low-metallicity environment of NGC 6822 with the goal of identifying nearby RSG populations and testing stellar evolution models by comparing them against our observational findings. 

NGC 6822, also known as Barnard's Galaxy, was discovered by \citet{Barnard1884} and is a barred irregular member of the Local Group located at a distance of around 0.5 Mpc \citep{vandenbergh2000}. Searching for RSGs in NGC 6822 was particularly appealing in part because of the recently available NIR photometric data from the United Kingdom Infra-Red Telescope (UKIRT), and parallax and proper motion values of its stars from the Gaia data releases \citep{GaiaDR2, GaiaDR3}. With these datasets, we can use the same techniques applied in the Local Group galaxies of M31, M33, LMC, SMC, and IC 10 to identify RSGs in NGC 6822. NGC 6822 also has a low metallicity of $0.4Z\odot$ (see further discussion in \citealt{EmilyNGC6822}) which will allow for future studies of the RSG content over a range of different metallicities.

NGC 6822 has been the subject of several previous RSG content studies including \citet{Massey1998}, \citet{EmilyNGC6822}, \citet{Patrick2015}, and \citet{Yang2021}. The largest sample was identified by \citet{Yang2021}, which identified 234 RSGs, but the survey suffered from a discrepancy with the photometry (see Section 3.2) that introduces problems in their selection criteria. Here we propose a sample of the $K$-band brightest 51 RSGs in NGC 6822 (many, but not all, of these RSGs are in common with those previously discovered) and additionally compare our observational findings to theoretical models of single and binary star evolution. 

In Section 2 we identify NGC 6822 RSGs after removing both foreground Galactic red dwarfs and asymptotic giant branch (AGB) contaminants from our initial sample. We additionally compare our final list of RSGs with results from previous work and discuss our overall completeness. In Section 3 we then determine their physical properties such as temperature and luminosity before comparing these observed properties to evolutionary model predictions in Section 4. Finally, in Section 5 we briefly summarize our research and discuss next steps.

\section{Identifying RSGs}
\subsection{NIR Photometry}
To locate RSGs in NGC 6822, we relied on the archival NIR dataset taken by \citet{Irwin2013} using the Wide Field Camera (WFCAM) on the 3.8-m UKIRT on Mauna Kea. These data were taken as part of a survey of the Local Group galaxies. We began with the pipeline reduced source catalogs in $J$ and $K$, and identified sources that were flagged as ``stellar” or ``marginally stellar” that appeared in both filters, following the same procedure used in our study of IC 10 \citep{Dimitrova2020} and  M33 \citep{Massey2021}. Comparison with 2MASS photometry \citep{2MASS} of the same stars led us to revise the UKIRT photometry by +0.023 mag in $J$ and -0.016 mag in $K$, consistent with our previous studies; thus the correction in $J-K$ colors are 0.039 mags. The comparison also confirmed that the UKIRT photometry became unreliable brighter than $J=K=12$.  We therefore added stars from the 2MASS catalog covering the same area.  All together, our initial source list consisted of 213,697 stars, of which 211,473 were identified from the UKIRT data, and the remainder from 2MASS.

We cross-matched this list with the Gaia DR2 \citep{GaiaDR2}, finding 115,106 (54\%) stars with complete Gaia data.  We will use this below to establish membership in NGC 6822.

\subsection{Removing Contaminants}
While the UKIRT NIR data allows us to identify red stars in the direction of NGC 6822, foreground Galactic  nearby red dwarfs (and the occasional halo red giant) fall into the same magnitude range as distant, but much more luminous, RSGs in NGC 6822. To remove the red foreground stars, we relied on high-precision parallax ($\pi$) and proper motion values ($\mu_\alpha$, $\mu_\delta$) from the Gaia Data Releases to assign membership in NGC 6822.  However, as is well known, there are spatially dependent zero-point issues with Gaia parallaxes and proper motions values, and in order to account for these, we used the same process described in \citet{LMCBinFrac} and \citet{Massey2021}, based upon the recommendations given in \citet{GaiaDR2}.  We first identified a set of stars that were near-certain members of NGC~6822.  For this, we selected probable OB stars from the Local Group Galaxy Survey (LGGS; \citealt{LGGSII}). A sample of blue stars was defined using the $B-V$, $U-B$, and $V$ bands, with values from $B-V < 0.1$, $U-B < -0.7$, and $V = 18$ to 20 based on examining spectroscopically confirmed OB stars in NGC 6822 \citep{LGGSII}. While our NIR photometric catalog was cross-matched against Gaia DR2, we relied on the newly released Early Gaia Data Release 3 (EDR3; \citealt{GaiaDR3}) to determine the zero-points, as this resulted in more accurately determined zero-point values. By averaging the blue star members' proper motion and parallax values, we determined the following Gaia zero points for NGC 6822 members: $\pi = 0.16\pm0.07$ mas, $\mu_\alpha = -0.30\pm0.11$ mas yr$^{-1}$, and $\mu_\delta = -0.39\pm0.14$ mas yr$^{-1}$, where the quoted errors are the standard deviations of the mean.  Note that all three values differ by multiple sigma from zero. We then used the Gaia DR2 proper motion and parallax values, along with their estimated uncertainties, to identify near-certain members of NGC 6822. For this, we assigned membership status if all three values were within three times their uncertainties of the expected zero-points. (We were careful not to exclude stars whose only problem was that their parallax values were too small compared to this criterion.)  In all, 20,729 (18\%) of the 115,106 with Gaia data met these criteria for membership.

\begin{figure}[ht!]
\centering
\includegraphics[width=0.5\columnwidth]{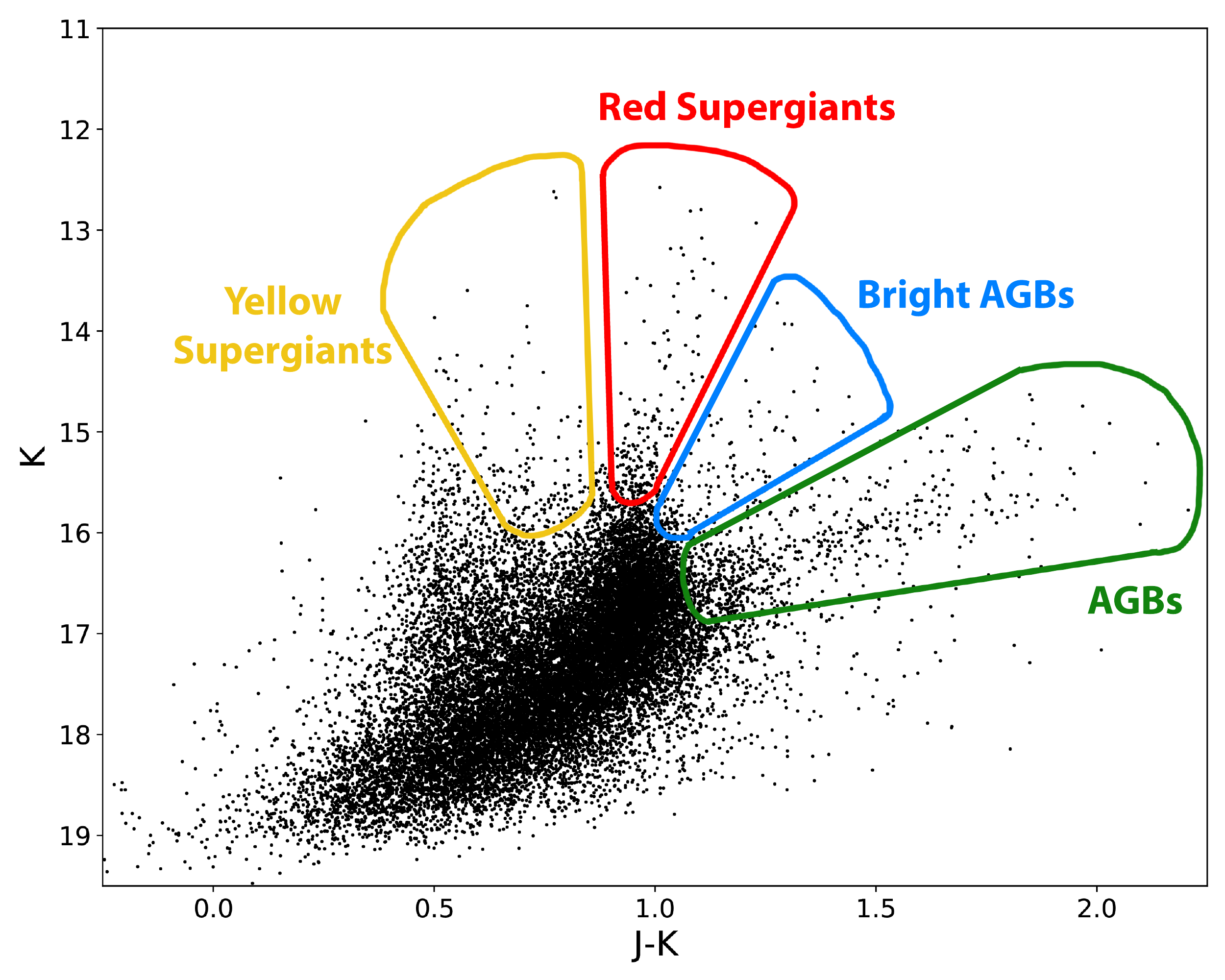}
\caption{CMD constructed from the $K$ and $J-K$ values of NGC 6822’s stellar members with foreground stars removed using Gaia. The approximate regions of different stellar populations are as labeled above, revealing the presence of both YSGs on the left side of the approximate RSGs on the CMD, and AGB stars to their right on the CMD.\label{fig:CMD}}
\end{figure}

After the removal of foreground stars, we next needed to remove non-RSG contaminants such as lower mass AGBs and cooler YSGs, also residing in NGC 6822, as shown in Figure~\ref{fig:CMD}. To do this, a color magnitude diagram (CMD) was constructed from the $J-K$ and $K$ values of the remaining sample. Originally, we attempted to apply $J-K$ and $K$ cuts used previously to identify RSGs in the LMC by \citet{LMCBinFrac} since the metallicity of the LMC ($0.5Z_\odot$, \citealt{RussellDopita}) is comparable to that of NGC 6822 ($0.4Z_\odot$). However, even after accounting for NGC 6822's higher reddening compared to the LMC ($E(B-V) = 0.25$ for NGC 6822 vs.\ 0.13 for the LMC; \citealt{LGGSII}) and distance (0.5 Mpc vs.\ 0.05 Mpc, respectively; \citealt{vandenbergh2000}), the cuts did not adequately select the RSGs and instead preferentially selected YSGs, suggesting our $J-K$ cutoffs were too low. Given this method worked with great success in both M31 and M33 \citep{Massey2021}, as well as in the SMC \citep{WRRSG} and IC 10 \citep{Dimitrova2020}, we do not have a sound explanation for why a transformation of the equations was not successful in NGC 6822. A few possibilities include difficulties with the higher reddening value, the slight metallicity difference between the LMC and NGC 6822, or problems arising from the small number of candidate RSGs in NGC 6822. 

\begin{figure}[ht!]
\centering
\includegraphics[width=0.5\columnwidth]{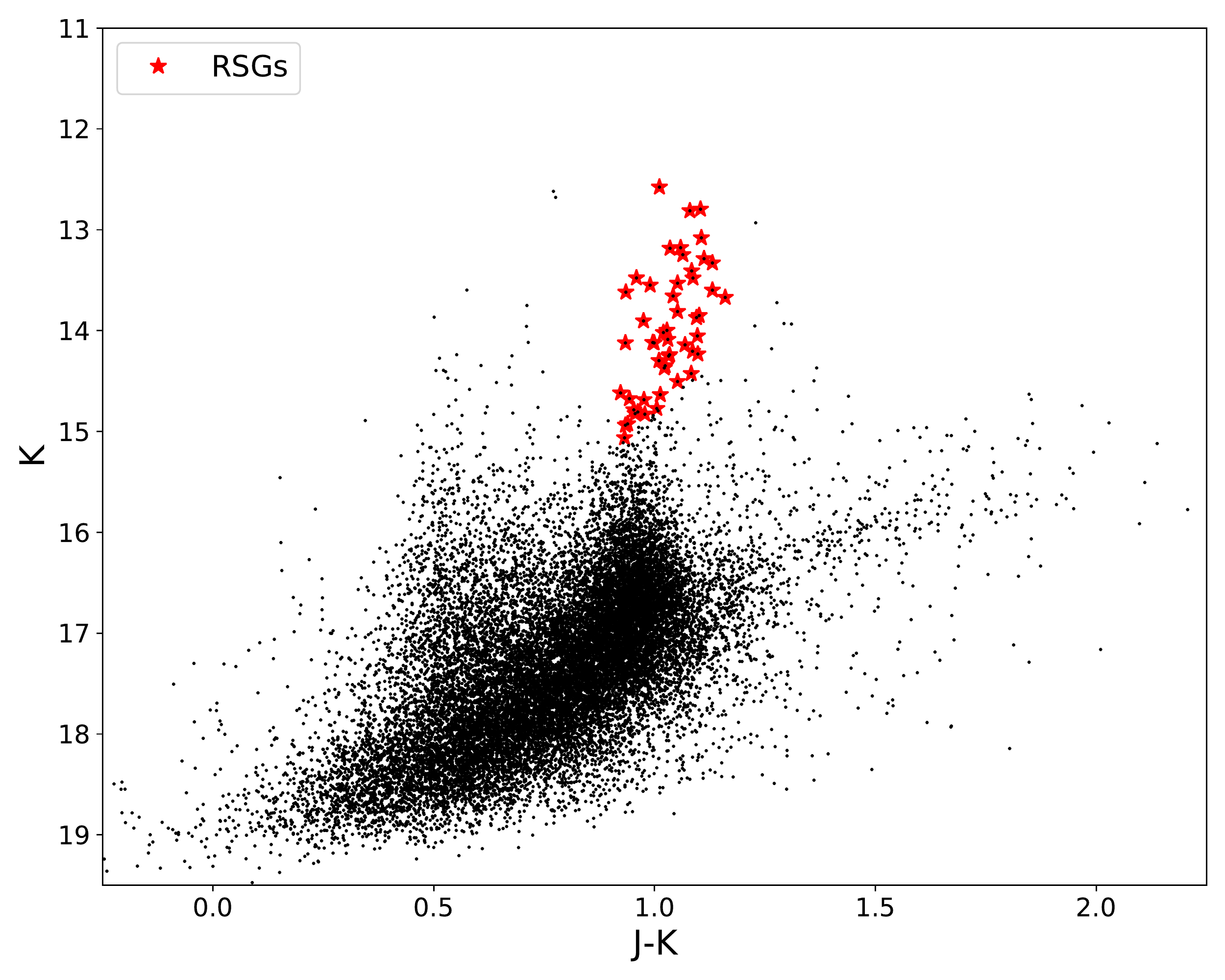}
\caption{CMD of the final 51 $K$-band brightest RSGs in NGC 6822. The RSGs are plotted in red and the galaxy's other stellar members are shown in black. The limits are defined in (J-K, K) space as: (0.9, 15.2), (1.2, 14), (0.9, 11), and (1.2, 11). \label{fig:CMDRSGs}}
\end{figure}

Instead of attempting to add an offset to the poorly-matched $J-K$ cuts, the photometric selections of the most luminous RSGs within NGC 6822 were instead made by eye. Due to our inability to use the methodology described above, our list of RSGs is not complete down to a limiting luminosity. However, our $J-K$ and $K$ selections were made to instead select the most luminous RSGs without running the risk of selecting any high-luminosity AGB stars or cooler YSGs. The selected 51 of the most $K$-band brightest RSGs in NGC 6822 are shown in Figure~\ref{fig:CMDRSGs} and their coordinates, magnitudes, and colors are listed in Table~\ref{tab:NGC6822RSGs}. 

\subsection{Comparative Studies and Completeness}
We have identified the 51 $K$-band brightest RSGs in NGC 6822. However, given our difficulties photometrically differentiating the RSGs from the YSGs and AGBs, we are not able to claim completeness down to a limiting luminosity. To get a better sense of our completeness, we can compare our sample to previous studies of RSGs in NGC 6822 to further test various steps of our method. 

\begin{figure}[ht!]
\centering
\includegraphics[width=0.5\columnwidth]{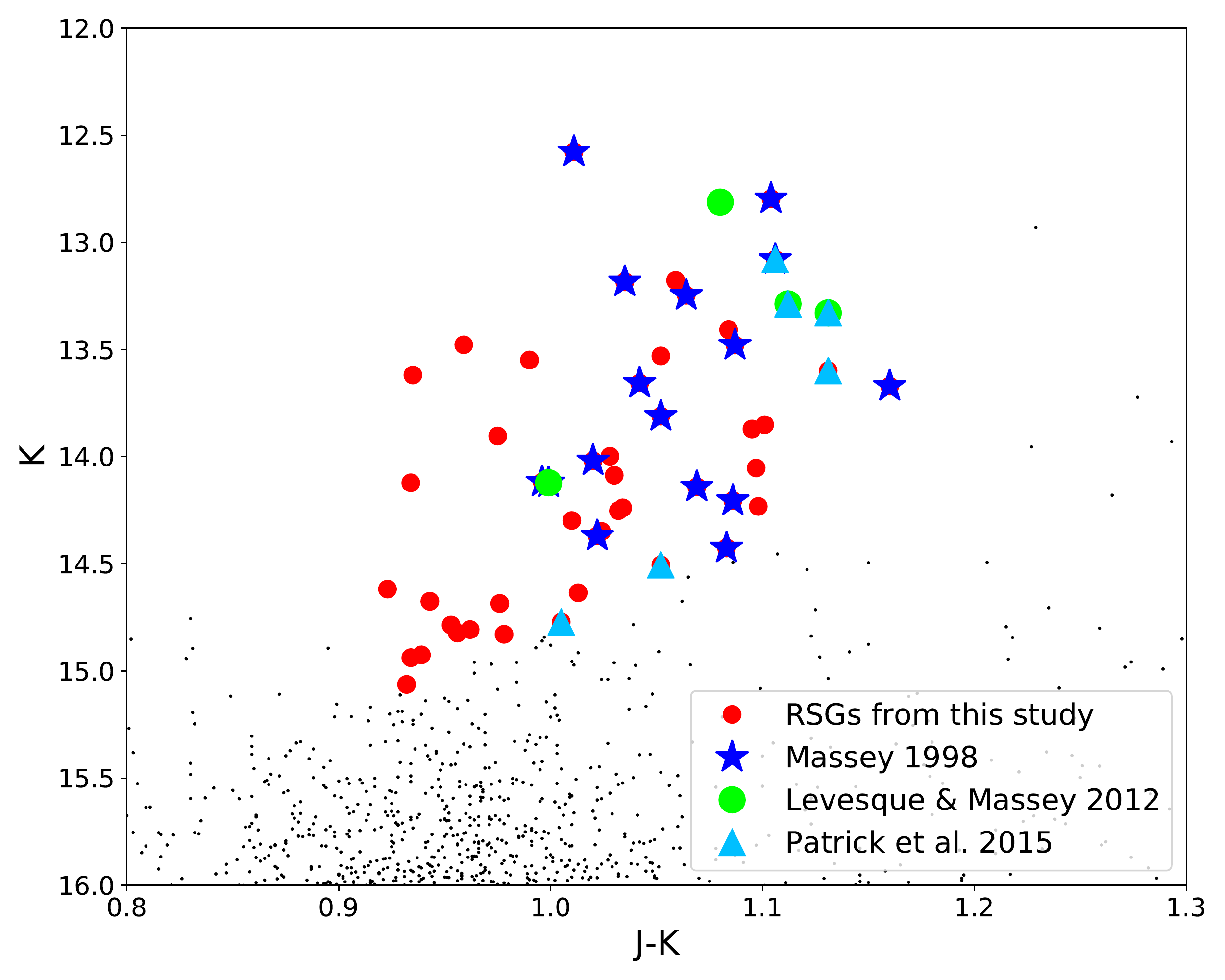}
\caption{Color Magnitude Diagram summarizing all of the sample matches with the
comparative data sets. The \citet{Massey1998} comparison is shown in dark blue stars and produced the most matches within our data set, with 16 matching RSGs. The \citet{Patrick2015} comparison is
shown in light blue triangles and contains 6 matching RSGs. The \citet{EmilyNGC6822} comparison is shown in green circles and contains 4 matching RSGs.\label{fig:otherStudies}}
\end{figure}

The first study of RSGs in NGC 6822 was done by \citet{Massey1998}, which probed the evolution of massive stars as a function of metallicity in various Local Group galaxies and investigated RSGs within a small subset of NGC 6822. This study was conducted prior to Gaia and contained both RSG candidates and foreground Galactic dwarfs that were then assigned membership through radial velocity measurements and the strength of the Ca II triplet. \citet{Massey1998} observed 199 candidate RSGs in NGC 6822 and classified 77 of them as RSGs and the remaining 122 stars as foreground red dwarfs. We first compared our list of NGC 6822 members to their list of 77 RSGs and found that 50 of them are indeed likely RSGs while the remaining 27 are foreground red dwarfs. Of the 122 foreground stars classified by \citet{Massey1998}, 16 of them are actually members based on the Gaia proper motion and parallax data. We then compared our list of 51 RSGs to their initial list of 199 stars and identified 19 matches. Of these, 16 of them were identified as RSGs by \citet{Massey1998} while the remaining 3 were identified as foreground stars. Of the remaining 34 RSGs identified by \citet{Massey1998} but not in our list of 51 RSGs, 32 of them were simply too faint with $K$ magnitudes dimmer than 15, and 2 are in the region of bright AGBs. The discrepancy in membership status between this survey and that of \citet{Massey1998} is due to his lack of Gaia data and reliance on radial velocities to define membership. The radial velocity of NGC 6822 varies across the disk of the galaxy between -40 km s$^{-1}$ and -90 km s$^{-1}$, which are also reasonable radial velocities for Galactic stars. Thus, this overlap can lead to confusion when determining membership. We plot the magnitudes of the stars that matched between our sample and that of \citet{Massey1998} over the CMD of our candidates in Figure~\ref{fig:otherStudies}.

We next compared our sample to the work done by \citet{EmilyNGC6822} which was also done before the release of Gaia and relied on membership determination through a combination of $B-V$ vs.\ $V-R$ color-cuts and radial velocity measurements. Their data contained a sample of 16 RSGs in NGC 6822 that were observed spectroscopically and analyzed to determine their spectral types. Of these 16 RSGs, only 8 were identified as members of NGC 6822 after being filtered through Gaia. Of the remaining 8, only 4 were within our list of 51. One of the remaining stars is fainter than our sample and the other three fall into the regime of bright AGBs. We plot the matching stars in Figure~\ref{fig:otherStudies}.

Our candidate sample was then compared to the spectroscopic observations of 18 RSGs in NGC 6822 by \citet{Patrick2015}. Of these 18 RSGs, 11 of them appear to be members based on the Gaia data while the remaining 7 are likely foreground red dwarfs (including two of the stars they used for abundance analysis). Of the remaining 11, 6 stars are in common with our list of 51 RSGs. The remaining 5 stars were not included in our sample because they were too dim in $K$. We again plot the matching stars in Figure~\ref{fig:otherStudies}.

We additionally attempted to compare our sample of RSGs to the most recent RSG study in NGC 6822 done by \citet{Yang2021} which also relied on the same archival UKIRT data. As part of their study, \citet{Yang2021} identified 234 RSGs in NGC 6822 using both the stars' $1.6\mu m$ ``H-bump" as well as color and magnitude cuts, and we attempted to use their $J-K$ and $K$ cuts to select our own sample of RSGs. However, after further investigation, we found that their $J$ magnitudes have a serious discrepancy with 2MASS. For the 42 stars in common with their RSG catalog and 2MASS, the $J$ magnitudes disagree by $\-0.23\pm0.01$ mag (in the sense of Yang-2MASS). Indeed, the issue can be seen within their own table when alternative sources for $J$-band photometry are included. There is no similar problem for their $K$-band photometry. This results in a large error in the $J-K$ colors, and thus their resulting CMD and selection criteria. Given this, we do not compare our sample to theirs. 

After comparing our list of 51 RSGs to the work done by previous surveys (a summary of which can be found in Table~\ref{tab:NGC6822RSGs}), we can draw a few conclusions both about previous studies as well as the completeness of our own. The first is that due to the overlap in the expected radial velocities between NGC 6822 and Galactic halo stars, radial velocities alone are not enough to determine membership in the galaxy. The second is that, while we have identified the brightest RSGs in NGC 6822, there are many fainter RSGs with $K$ magnitudes dimmer than 15th magnitude that likely still exist. However, given our inability to separate these RSGs from the bright AGBs, we cannot use this method to identify them. Finally, separating RSGs from bright AGBs, even with high resolution spectra, is still an ongoing area of research. While we have classified a few of the stars observed by \citet{EmilyNGC6822} and \citet{Patrick2015} as bright AGBs rather than RSGs, their evolutionary status is still an open question. Overall, we emphasize that this study lacks completeness but instead presents a survey of the brightest RSGs in NGC 6822.

\section{Determining Temperatures and Luminosities}
The observed physical properties of NGC 6822's RSGs can ultimately be compared to the theoretical work of stellar evolutionary models by first determining their effective temperatures and luminosities. Starting with the $K$ and $J-K$ UKIRT photometry, these colors and magnitudes were transformed to the Bessell \& Brett photometric system \citep{Bessell98} using the following equations from \citet{Carpenter}: 
\[K_B = K + 0.044\]
\[(J-K)_B = (J-K + 0.11) / 0.972\]
Where $(J-K)_B$ and $K_B$ are the converted $K$ and $J-K$ bands in this photometric system. Next the photometry was de-reddened using $E(B-V) = 0.25$ as was determined for NGC 6822 by \citet{LGGSII}. The $K$ and $J-K$ photometry underwent de-reddening transformations using the reddening relations derived from \citet{1998ApJ...500..525S} as given in \citet{UKIRTpap}:
\[E(B-V) = 0.25\]
\[K_0= K_B - 0.367 \times E(B-V)\]
\[(J-K)_0 = (J-K)_B - 0.54 \times E(B-V)\]
With $K_0$ and $(J-K)_0$ being the de-reddened photometry. We then computed the absolute $K$ magnitudes corresponding to a true distance modulus of 23.49 based upon an adopted distance to NGC 6822 of 0.50~Mpc \citep{vandenbergh2000}.

The transformed and de-reddened $J-K$ band values were then used to calculate the effective temperatures of the stars in NGC 6822, following the equation from \citet{LMCBinFrac} with values adjusted for NGC 6822's distance: 
\[T_{\rm{eff}} = 5643.5 - (1807.1 \times (J-K)_0)\]
Next, the bolometric correction was determined from the effective temperature values calculated above, to convert the stars’ K magnitudes to bolometric magnitudes. This was done with an equation similar to that in \citet{LMCBinFrac}.
\[BC_K = 5.567 - (7.5686 \times 10-4) \times T_{\rm{eff}}\]
Where $BC_K$ is the bolometric correction. The bolometric corrections were then applied and a bolometric magnitude was used to determine the luminosity values of the stars in NGC 6822. As described in \citet{LMCBinFrac}, uncertainties in $\log T_{\rm{eff}}$ and $\log\frac{L}{L_\odot}$ are 0.02 dex and 0.05 dex, respectively.

We list the derived temperatures and luminosities of our 51 RSGs in Table~\ref{tab:NGC6822RSGs} and compare these values to the theoretical predictions of binary and single star evolutionary models in the next section.

\section{Comparing our Observations with Evolutionary Model Predictions}
We compared our observations with theoretical predictions from two different stellar evolutionary codes: the Geneva evolutionary models \citep{GenevaLMC}, which focus on single-star evolution as well as the Binary Population and Spectral Synthesis (BPASS) models \citep{BPASS1, BPASS2}, which include both single and binary star evolution. For both of these evolutionary codes, we used the LMC metallicity ($z = 0.006$) models. 
 
\begin{figure}[ht!]
\centering
\includegraphics[width=0.5\columnwidth]{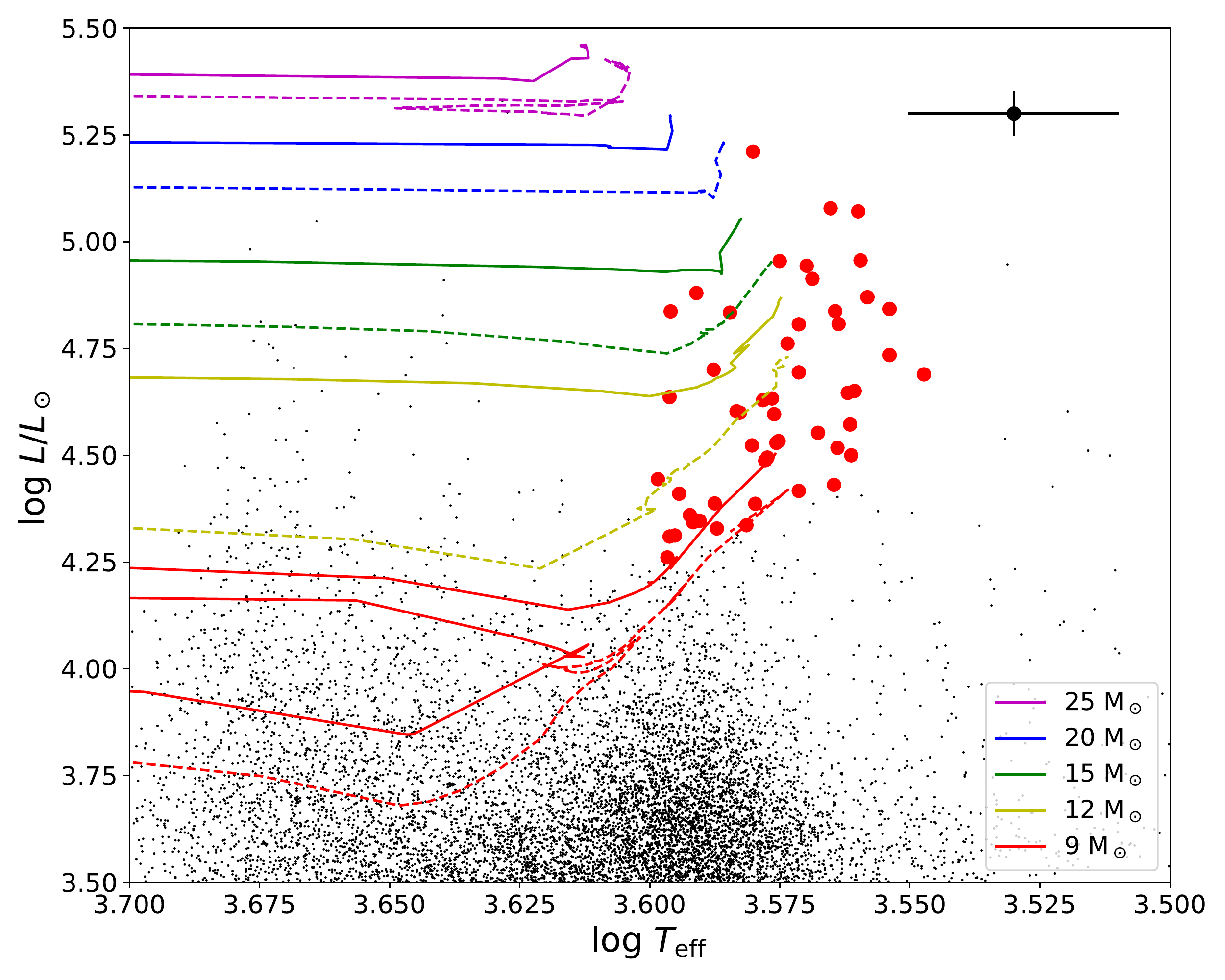}
\caption{HR Diagram showing the Geneva evolutionary tracks compared to our proposed RSGs in red, and the remainder of NGC 6822's stellar population in black. Uncertainty values for all points are shown in the upper right corner with further discussion in the text. The tracks displayed represent a stellar mass range of 9 to 25$M_\odot$ for $z = 0.006$ \citep{GenevaLMC}. Rotating models are shown in solid lines and non-rotating models are shown in dashed lines. The locations of the tracks generally fit the observed distribution of our RSGs well and reveal our sample of RSGs to be within a mass range of 9 to 20$M_\odot$. Note, the group of RSGs located in the upper-right that fall outside these tracks, which is discussed in greater detail below.\label{fig:GenevaHRD}}
\end{figure}

A comparison between our observations and the Geneva models for stars with initial masses between $9-25M_\odot$ is shown in the HRD in Figure~\ref{fig:GenevaHRD}. As expected, the majority of the identified RSGs fall along the upturn at the cool edge of each evolutionary path, otherwise known as the Hayashi track. These tracks additionally suggest that the brightest RSGs in NGC 6822 have initial masses between $9 - 20M_\odot$. This lower limit of $9 M_\odot$ suggests that we have done an adequate job selecting only the more massive RSGs as opposed to the lower mass AGBs which would come from stars with initial masses below $9 M_\odot$. For the upper limit, while the Geneva models predict that RSGs can form from stars with initial solar masses as high as $30M_\odot$ at LMC-like metallicities, we do not find any of these more massive RSGs in NGC 6822. This is likely due to the small number of RSGs present in NGC 6822, as the lifetimes of the RSG stage will be shorter with increasing mass. It is additionally important to note that some of our observed RSGs extend to cooler temperatures than what the Geneva models predict for the metallicity regime. An explanation for this small disagreement between the observations and the models is offered below.

\begin{figure}[ht!]
\includegraphics[width=0.5\columnwidth]{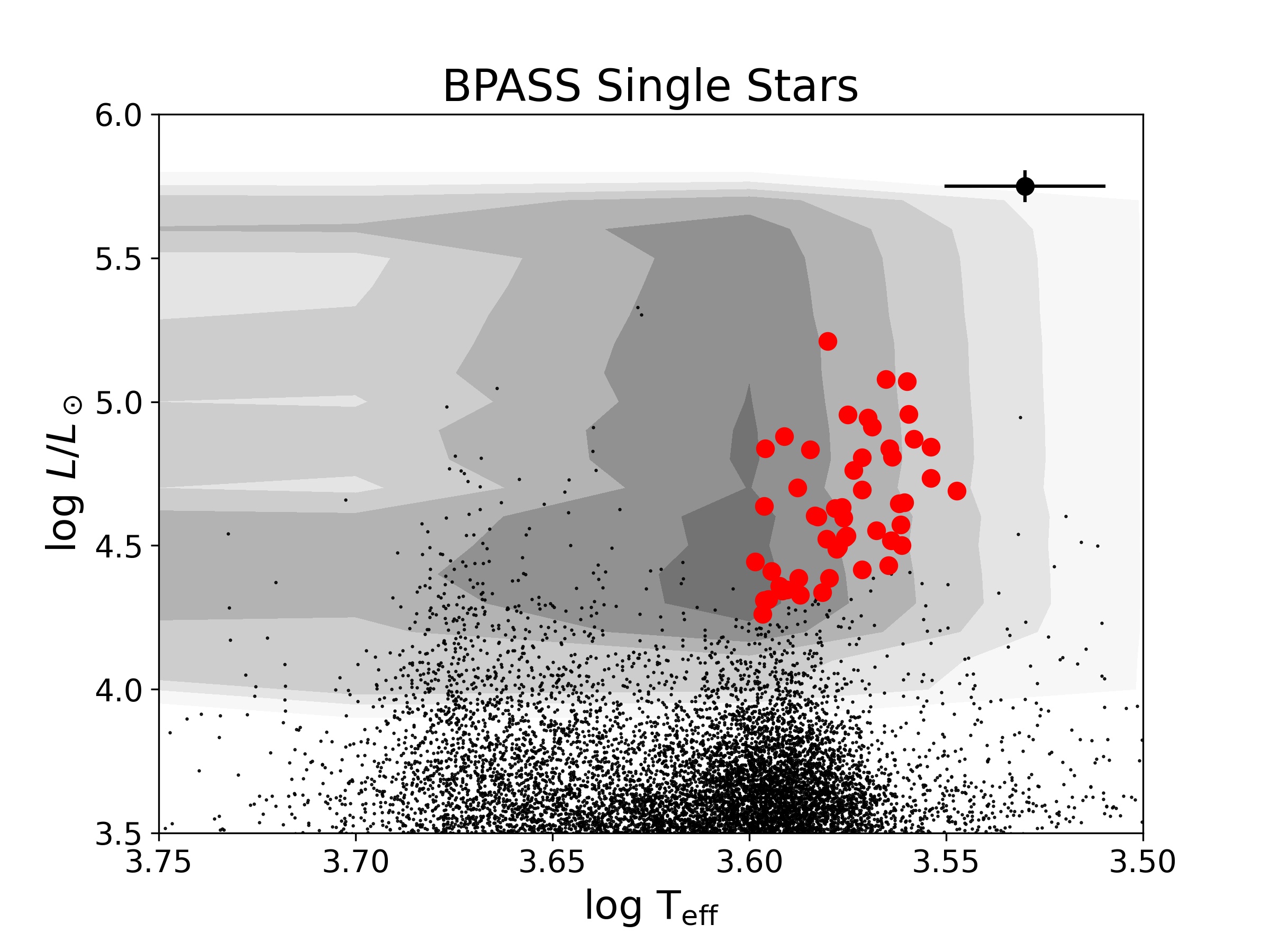}
\includegraphics[width=0.5\columnwidth]{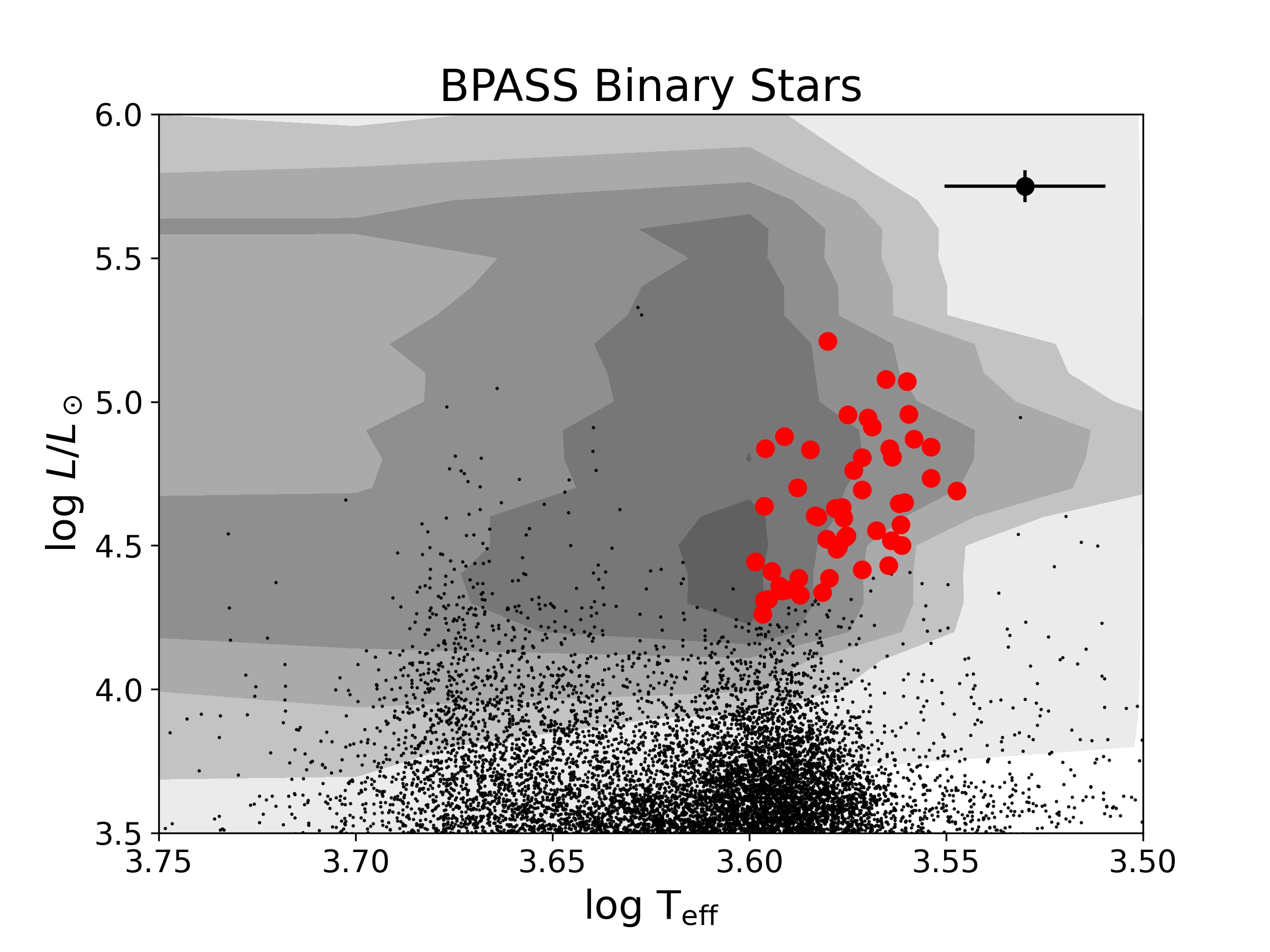}
\caption{HR Diagrams comparing the BPASS single (left) and binary (right) compared to our proposed RSGs in red, and the remainder of NGC 6822's stellar population in black. Uncertainty values for all points are shown in the upper right corner with further discussion in the text. In both diagrams, population density is indicated by the shaded regions (darker colors indicate a denser population, and lighter colors indicate a less dense population). Note the agreement between the BPASS single and Geneva evolutionary tracks but the improved agreement between our observations and the models once binaries are included in the BPASS models. This BPASS model that includes binaries better models the ``bump" of cooler RSGs and follows the overall distribution pattern more accurately. This BPASS model for a mixture of single and binary stars is the best-fit model for our proposed RSGs, thereby suggesting the importance of binary interactions.} \label{fig:BPASS}
\end{figure}

The locations of the BPASS evolutionary tracks were also compared to the physical properties of our final candidates for both single and binary star evolution, as is shown in the HRD in Figure~\ref{fig:BPASS}. These BPASS models produce density maps indicating predictions of effective temperature and luminosity positions on the HRD through a stellar lifetime, with shaded regions corresponding to theoretical population density. The single star evolution model shows a very similar prediction to the Geneva tracks shown in Figure~\ref{fig:GenevaHRD}, partially excluding the region of RSGs that skew to the right. Interestingly, the BPASS model for a mixture of single and binary RSGs predicts a bump of RSGs to the right, aligning with our population of cooler RSGs. This suggests that our sample includes either current or previous binary RSGs, which would theoretically affect the RSG distribution. 

One way to determine whether the cooler stars represent current RSG binaries or just the products of past binary interactions is to compare the HRD shown in Figure~\ref{fig:BPASS} with the HRD shown in Figure 6 of \citet{LMCBinFrac}. Their HRD of RSGs in the LMC is color-coded by the probability of binarity for each star with blue dots representing binary RSGs and red dots representing single RSGs. Note that the binary RSGs are \emph{not} clustered in a singular area of the HRD and thus the presence of a binary companion does not appear to be correlated with the temperature of the RSG. This suggests that this cooler population of RSGs we observe in the HRD of NGC 6822 is not due to the presence of currently binary RSGs. Instead, it is perhaps the end result caused by previous stages of binary evolution. 

An alternate explanation to this cooler population of RSGs is that higher luminosity RSGs are known to be both more variable and also have higher mass-loss rates than lower luminosity RSGs \citep{Meynet2015, Humphreys2020}. It is possible that these high luminosity RSGs have recently gone through eruptive mass-loss events similar to the recent outburst by Betelgeuse \citep{EmilyBetelgeuse, BetelgeuseNature} and thus appear cooler since much of their light has been blocked by dust. However, as discussed briefly by \citet{WRRSG}, when a similar group of cooler RSGs was discovered in the SMC, spectroscopy confirmed them to be cool M-type RSGs as opposed to highly reddened and dusty K-type stars. While it is impossible to differentiate any of these effects from the intrinsic uncertainties in our temperature calculations, the good agreement between our observations and the BPASS binary evolutionary models is still striking even if alternative explanations for this cooler population of RSGs exist.

\section{Summary, Conclusions, and Next Steps}
We present a list of the 51 $K$-band brightest RSGs in NGC 6822. These RSGs were discovered after using Gaia parallax and proper motion values to determine membership, archival NIR photometry from UKIRT to constrain stellar characteristics, and $K$ and $J-K$ cut-offs from the constructed CMD to distinguish RSGs from AGBs. This sample was studied comparatively alongside several other RSG searches in NGC 6822, revealing our results to be in general agreement with these previous studies and additionally discovering that radial velocity is not an adequate method of determining membership in NGC 6822 due to Galactic halo star contamination. Our RSGs were compared with the Geneva and BPASS stellar evolutionary models for LMC-like metallicities as a means of testing theoretical predictions to observational findings in the galaxy, confirming the models as a good fit for the RSGs in NGC 6822 and in particular identifying the BPASS model for a mixture of binary and single stars as the best-fit scenario. This result underscores the importance of binary interactions both before and during the RSG phase when studying the RSG population of a galaxy. 

This proposed RSG sample has broader applications for the overall study of RSG populations in the Local Group, as it provides a low-metallicity complement to nearby galaxies in which previous RSG searches have occurred. This research is also complementary to our previous work, in which we located RSGs in the starburst galaxy IC 10. Future work comparing our results in IC 10 and NGC 6822 with the RSG searches done by our collaborators in M31, M33, and the Magellanic Clouds, could be beneficial to further testing the existing stellar evolution models and theoretical predictions. Additionally, now that a new population of RSGs in NGC 6822 has been located, future investigations into these massive stars with direct spectroscopic follow-up, and studies into the binary frequency of the RSGs in this low metallicity galaxy, are supported. Furthermore, our work reveals an interesting group of cool RSGs, potentially indicating the presence of RSGs effected by binary evolution and creating an opportunity for future RSG binary related probes within this and other galaxies.

\acknowledgments
We thank both Ming Yang and the anonymous referee for their helpful comments that improved the paper. TAD would like to acknowledge financial support from both the William Goddard Family through Lowell Observatory and the Mary Gates Foundation through the University of Washington. This work was additionally supported in part by the Dunlap Institute for KFN, the National Science Foundation under grant AST-1612874 awarded to PM, and a Cottrell Scholar Award from the Research Corporation for Scientific Advancement granted to EML.

\begin{deluxetable*}{l l l l l l l l l l}
\tabletypesize{\scriptsize}
\tablecaption{\label{tab:NGC6822RSGs} $K$-Band Brightest RSGs in NGC 6822}
\tablewidth{0pt}
\tablehead{\colhead{$\alpha_{\rm 2000}$} 
& \colhead{$\delta_{\rm 2000}$} 
& \colhead{$K$}
& \colhead{$\sigma K$}
& \colhead{$J-K$}
& \colhead{$\sigma J-K$}
& \colhead{$\log T_{\rm{eff}}$}
& \colhead{$\log L/L_\odot$}
& \colhead{Other Survey\tablenotemark{*}}
}
\startdata
19 42 21.59 & -14 45 11.9 & 14.122 & 0.005 & 0.934 & 0.009 & 3.60 & 4.64 & \\
19 43 30.67 & -14 32 35.7 & 14.807 & 0.007 & 0.962 & 0.011 & 3.59 & 4.35 & \\
19 44 05.19 & -14 11 25.0 & 13.619 & 0.003 & 0.935 & 0.005 & 3.60 & 4.84 & \\
19 44 29.74 & -14 52 36.9 & 14.635 & 0.007 & 1.013 & 0.011 & 3.58 & 4.39 & \\
19 44 30.35 & -14 48 35.9 & 13.549 & 0.004 & 0.990 & 0.006 & 3.58 & 4.83 & \\
19 44 32.09 & -14 49 26.7 & 13.178 & 0.003 & 1.059 & 0.005 & 3.57 & 4.94 & \\
19 44 35.29 & -14 40 45.7 & 14.426 & 0.005 & 1.083 & 0.009 & 3.56 & 4.43 & 1 \\
19 44 43.27 & -14 51 51.5 & 14.786 & 0.007 & 0.953 & 0.011 & 3.59 & 4.36 & \\
19 44 43.80 & -14 46 11.0 & 13.080 & 0.003 & 1.106 & 0.005 & 3.56 & 4.96 & 1,3 \\
19 44 43.83 & -14 51 31.3 & 14.232 & 0.005 & 1.098 & 0.009 & 3.56 & 4.50 & \\
19 44 45.04 & -14 51 18.6 & 13.530 & 0.003 & 1.052 & 0.006 & 3.57 & 4.81 & \\
19 44 45.76 & -14 52 21.7 & 12.812 & 0.002 & 1.080 & 0.004 & 3.57 & 5.08 & 2 \\
19 44 45.80 & -14 51 08.1 & 14.829 & 0.008 & 0.978 & 0.012 & 3.59 & 4.33 & \\
19 44 45.98 & -14 51 02.8 & 14.505 & 0.006 & 1.052 & 0.010 & 3.57 & 4.42 & 3 \\
19 44 47.13 & -14 35 06.9 & 13.871 & 0.004 & 1.095 & 0.006 & 3.56 & 4.65 & \\
19 44 47.80 & -14 50 52.9 & 13.287 & 0.003 & 1.112 & 0.005 & 3.56 & 4.87 & 2,3 \\
19 44 48.09 & -14 45 18.4 & 14.019 & 0.005 & 1.020 & 0.008 & 3.58 & 4.63 & 1 \\
19 44 50.42 & -14 44 10.4 & 14.122 & 0.005 & 0.999 & 0.008 & 3.58 & 4.60 & 1,2 \\
19 44 51.09 & -14 43 55.8 & 12.577 & 0.002 & 1.011 & 0.004 & 3.58 & 5.21 & 1 \\
19 44 51.56 & -14 43 20.5 & 13.479 & 0.002 & 1.087 & 0.004 & 3.56 & 4.81 & 1 \\
19 44 53.36 & -14 45 24.7 & 14.141 & 0.005 & 1.069 & 0.009 & 3.57 & 4.55 & 1 \\
19 44 54.17 & -14 53 04.1 & 14.239 & 0.005 & 1.034 & 0.009 & 3.58 & 4.53 & \\
19 44 54.24 & -14 49 35.0 & 14.053 & 0.064 & 1.097 & 0.084 & 3.56 & 4.57 & \\
19 44 54.46 & -14 48 06.5 & 13.329 & 0.003 & 1.131 & 0.005 & 3.55 & 4.84 & 2,3 \\
19 44 54.71 & -14 52 46.9 & 14.298 & 0.005 & 1.010 & 0.009 & 3.58 & 4.52 & \\
19 44 54.80 & -14 43 48.1 & 12.796 & 0.002 & 1.104 & 0.004 & 3.56 & 5.07 & 1 \\
19 44 54.86 & -14 52 16.7 & 14.252 & 0.005 & 1.032 & 0.009 & 3.58 & 4.53 & \\
19 44 55.77 & -14 45 30.2 & 13.671 & 0.004 & 1.160 & 0.006 & 3.55 & 4.69 & 1 \\
19 44 56.14 & -14 43 21.0 & 14.370 & 0.004 & 1.022 & 0.006 & 3.58 & 4.49 & 1 \\
19 44 56.73 & -14 44 25.5 & 13.811 & 0.004 & 1.052 & 0.006 & 3.57 & 4.69 & 1 \\
19 44 57.39 & -14 54 08.4 & 13.998 & 0.005 & 1.028 & 0.008 & 3.58 & 4.63 & \\
19 44 57.61 & -14 48 28.3 & 13.904 & 0.004 & 0.975 & 0.006 & 3.59 & 4.70 & \\
19 44 57.70 & -14 49 17.3 & 14.349 & 0.006 & 1.024 & 0.009 & 3.58 & 4.49 & \\
19 44 57.97 & -14 44 09.4 & 14.118 & 0.005 & 0.996 & 0.008 & 3.58 & 4.60 & 1 \\
19 44 58.16 & -14 44 56.4 & 13.657 & 0.004 & 1.042 & 0.006 & 3.57 & 4.76 & 1 \\
19 44 58.25 & -14 45 51.0 & 13.247 & 0.003 & 1.064 & 0.005 & 3.57 & 4.91 & 1 \\
19 44 58.31 & -14 44 47.2 & 13.184 & 0.003 & 1.035 & 0.005 & 3.58 & 4.95 & 1 \\
19 44 59.41 & -14 42 17.9 & 14.205 & 0.005 & 1.086 & 0.008 & 3.56 & 4.52 & 1 \\
19 44 59.56 & -14 52 30.6 & 14.685 & 0.007 & 0.976 & 0.011 & 3.59 & 4.39 & \\
19 45 00.32 & -14 41 05.0 & 13.851 & 0.004 & 1.101 & 0.006 & 3.56 & 4.65 & \\
19 45 00.54 & -14 48 26.8 & 13.599 & 0.004 & 1.131 & 0.006 & 3.55 & 4.73 & 3 \\
19 45 01.12 & -14 54 38.1 & 14.938 & 0.008 & 0.934 & 0.013 & 3.60 & 4.31 & \\
19 45 03.10 & -14 32 03.6 & 14.618 & 0.006 & 0.923 & 0.009 & 3.60 & 4.44 & \\
19 45 05.74 & -14 54 58.1 & 14.087 & 0.005 & 1.030 & 0.008 & 3.58 & 4.60 & \\
19 45 05.83 & -14 55 17.7 & 14.925 & 0.008 & 0.939 & 0.013 & 3.60 & 4.31 & \\
19 45 06.97 & -14 50 31.6 & 14.772 & 0.007 & 1.005 & 0.011 & 3.58 & 4.34 & 3 \\
19 45 08.21 & -14 55 17.0 & 13.408 & 0.003 & 1.084 & 0.005 & 3.56 & 4.84 & \\
19 45 09.54 & -14 53 45.0 & 14.823 & 0.008 & 0.956 & 0.012 & 3.59 & 4.34 & \\
19 45 13.20 & -14 58 43.9 & 13.478 & 0.004 & 0.959 & 0.006 & 3.59 & 4.88 & \\
19 45 21.23 & -14 59 05.0 & 14.675 & 0.007 & 0.943 & 0.011 & 3.59 & 4.41 & \\
19 45 47.99 & -14 41 52.7 & 15.063 & 0.008 & 0.932 & 0.013 & 3.60 & 4.26 & \\
\enddata
\tablenotetext{*}{1 = \citealt{Massey1998}; 2 = \citealt{EmilyNGC6822}; 3 = \citealt{Patrick2015}.}
\end{deluxetable*}

\bibliography{bib}
\bibliographystyle{aasjournal}

\end{document}